**A New Homologous Series of Iron Pnictide Oxide Superconductors**

**$(Fe_2As_2)(Ca_{n+2}(Al,Ti)_nO_y)[n = 2,3,4]$**

Hiraku Ogino[1,3*], Kenji Machida[1,3], Akiyasu Yamamoto[1,3], Kohji Kishio[1,3], Jun-ichi Shimoyama[1,3], Tetsuya Tohei[2], and Yuichi Ikuhara[2]

[1]Department of Applied Chemistry, The University of Tokyo, 7-3-1 Hongo, Bunkyo-ku, Tokyo 113-8656, Japan

[2]Institute of Engineering Innovation, The University of Tokyo, 2-11-16 Yayoi, Bunkyo-ku, Tokyo 113-8656, Japan

[3]JST-TRIP, Sanban-cho, Chiyoda-ku, Tokyo 102-0075, Japan

e-mail: tuogino@mail.ecc.u-tokyo.ac.jp

## Abstract

We have discovered a new homologous series of iron pnictide oxides $(Fe_2As_2)(Ca_{n+2}(Al,Ti)_nO_y)[n = 2,3,4]$. These compounds have perovskite-like blocking layers between $Fe_2As_2$ layers. The structure of new compounds are tetragonal with space groups of *P4/nmm* for $n$ = 2 and 4 and *P4mm* for $n$ = 3, which are similar to those of $(Fe_2As_2)(Ca_{n+1}(Sc,Ti)_nO_y)[n = 3,4,5]$ found in our previous study. Compounds with $n$ = 3 and 4 have new crystal structures with 3 and 4 sheets of perovskite layers, respectively, including a rock salt layer in each blocking layer. The *a*-axis lengths of the three compounds are approximately 3.8 Å, which are close to those of FeSe and LiFeAs. $(Fe_2As_2)(Ca_6(Al,Ti)_4O_y)$ exhibited bulk superconductivity in magnetization measurement with $T_{c(onset)}$~36 K and resistivity drop was observed at ~39 K. $(Fe_2As_2)(Ca_5(Al,Ti)_3O_y)$ also showed large diamagnetism at low temperatures. These new compounds indicate considerable rooms are still remaining for new superconductors in layered iron pnictides.

## Introduction

After the discovery of high-temperature superconductivity in *RE*FeAsO family[1], several related compounds having antifluorite $Fe_2Pn_2$($Pn$ = pnictogen) layers has been developed. Among them, a new family of layered iron pnictides with perovskite-type oxide blocking layers has recently been discovered[2-12]. Since the perovskite-type layers are flexible in terms of chemical composition and crystal structure, there is large possibility to create new compounds by modifying the perovskite-type layers. In fact, many compounds, such as $(Fe_2As_2)(Sr_4M_2O_6)$ with $M$ = Sc, Cr, V and (Mg,Ti), were reported. The second bracket in each chemical formula represents the local chemical composition of the perovskite-type oxide layer. Moreover, the number of sheets in the perovskite-type layer can be controlled by the nominal compositions and the synthesis conditions as in the case of $(Fe_2As_2)(Ca_{n+1}(Sc,Ti)_nO_y)$ [$n$ = 3,4,5 and $y$ ~3$n$-1], which are discovered in our previous study[10]. These new compounds indicated the conditions for perovskite layer being suitable to form layered iron pnictide oxides, which are appropriate cell size with $a$ ~ 4 Å and having ability of charge compensation with the $(Fe_2As_2)^{-2}$ layer. In addition, the perovskite-type layers are divided into two types, $(AE_{n+1}M_nO_y)$ ($y$ ~3$n$-1, $AE$ = alkali earth metals) and $(AE_4M_2O_6)$ having a rock salt layer. The latter one gave us a hint to design new series, $(Fe_2As_2)(AE_{n+2}M_nO_y)$ ($y$ ~3$n$). Through the various attempts, we have discovered second homologous series of iron arsenide oxide superconductors $(Fe_2As_2)(Ca_{n+2}(Al,Ti)_nO_y)$[$n$ = 2,3,4, $y$ ~ 3$n$] in the present study.

**Experimental**

Samples with nominal compositions of $(Fe_2As_2)(Ca_{n+2}(Al_{1-x}Ti_x)_nO_y)$ [$n$ = 2,3,4] were synthesized by solid-state reaction from starting materials of FeAs (3N), Ca (2N), CaO (2N), Ti (3N), $TiO_2$ (3N) and $Al_2O_3$ (5N). Since the starting reagents were sensitive to moisture, manipulation was carried out in a glove box filled with argon gas. Powder mixtures were pelletized, sealed in evacuated quartz ampoules and heated at 1000-1200°C for 60-100 h followed by slow cooling to room temperature.

Constituent phases and lattice constants of resulting samples were analyzed by powder X-ray diffraction (XRD) measurements using Cu-$K_\alpha$ radiation (Rigaku Ultima-IV) and intensity data were collected in the $2\theta$ range of 5-80° at a step of 0.02°. Silicon powder was used as an internal standard. High angle annular dark field (HAADF) images were taken by scanning transmission electron microscope (STEM, Cs-corrected JEM-2100F, JEOL). TEM samples were prepared by crushing using agate pestle and mortar. Magnetic susceptibility was measured by a superconducting quantum interference device (SQUID) magnetometer (Quantum Design MPMS-XL5s).

Electrical resistivity was evaluated by the AC four-point-probe method using Quantum Design PPMS.

**Result and discussion**

Sintered bulk samples containing $(Fe_2As_2)(Ca_{n+2}(Al,Ti)_nO_y)[n = 2,3,4]$ as main phases were successfully obtained by optimizations of starting compositions and sintering conditions. Figure 1 shows powder XRD patterns of samples with nominal compositions of $(Fe_2As_2)(Ca_4(Al_{0.67}Ti_{0.33})_2O_6)$, $(Fe_2As_2)(Ca_5(Al_{0.5}Ti_{0.5})_3O_8)$ and $(Fe_2As_2)(Ca_6(Al_{0.33}Ti_{0.67})_4O_{11})$ reacted at 1000ºC, 1050 ºC and 1100 ºC, respectively, for 100 h. Crystal structure of $(Fe_2As_2)(Ca_4(Al,Ti)_2O_6)$ (abbreviated as 22426), $(Fe_2As_2)(Ca_5(Al,Ti)_3O_9)$ (22539) and $(Fe_2As_2)(Ca_6(Al,Ti)_4O_{12})$ (2264$\underline{12}$) are shown in Fig. 2. The structures of the compounds are tetragonal with space group *P4/nmm* for 22426 ($n$=2) and 2264$\underline{12}$ ($n$=4) and *P4mm* for 22539 ($n$=3) phases. Note that the number of perovskite cells, $n$, was increased up to 4 similar to $(Fe_2As_2)(Ca_{n+1}(Sc,Ti)_nO_y)$, whereas these compounds having a rock salt layer in each blocking layer.

As shown in Fig. 1(a), the 22426 phase formed as a main phase with several impurities, such as 22539, FeAs and $CaFe_2As_2$. It should be noted that the relative intensities of peaks due to impurity phases were minimum in a sample starting from $x$ = 0.33 and they monotonically increased either by increasing or by decreasing $x$ from 0.33. The sample contained ~60% of 22539 phase compared to 22426 phase estimated from intensities of their main peaks. The formation rate of 22539 phase was found to increase with increases in sintering temperature and/or titanium composition $x$, while the starting composition was $(Fe_2As_2)(Ca_4(Al_{1-x}Ti_x)_2O_6)$. On the other hand, lower sintering temperature or decreasing $x$ reduced the amount of 22539 phase in the resulting samples, however, phase purity of the 22426 phase were not improved due to generation of large amount of impurities, such as FeAs and $CaTiO_3$. The mean valence of $Al_{1-x}Ti_x$ in $(Fe_2As_2)(Ca_4(Al_{1-x}Ti_x)_2O_6)$ is calculated to be +3 and, hence, a starting composition with $x$=0 seems suitable to form the 22426 phase assuming that all titanium ions are tetravalent. However, any traces suggesting generation of iron pnictide oxides were not observed in XRD patterns of the samples with a nominal composition of $(Fe_2As_2)(Ca_4Al_2O_y)$. Further investigation will be needed to clarify the role of titanium doping and actual valence of titanium ions in this system. On the other hand, it is very recently reported that Ti-free $(Fe_2Pn_2)(Ca_4Al_2O_6)$ (Pn=As, P) are synthesized using high-pressure technique[13]. Possible reason for the absence of $(Fe_2As_2)(Ca_4Al_2O_6)$ phase in the present study is too small lattice size of $(Ca_4Al_2O_6)$ layer to stack with the $Fe_2As_2$ layer under ambient pressure. Although $a$-axis length of 22426 phase was not precisely analyzed due to broad XRD peaks, it was roughly estimated to be 3.77 Å. This value is similar to those of FeSe and LiFeAs.

Similar to the synthesis of 22426 phase, samples with nominal compositions of $(Fe_2As_2)(Ca_5(Al_{1-x}Ti_x)_3O_8)$ were composed of the 22539 phase as a major one and relatively large amounts of impurities, such as 226412 phase and FeAs. For example, intensity of the main peak of 226412 phase was almost half of that for the 22539 phase as shown in Fig. 1(b). On the other hand, the 226412 phase was obtained almost as single phase as shown in Fig. 1(c). There were no traces of 22426 and 22539 phases in the XRD pattern of 226412.

Comparing lattice constants of these new compounds, the *c*-axis lengths are systematically increased by ~4 Å with increasing *n*, corresponding to the unit-cell size of the perovskite. At the same time, *a*-axis length increased by ~0.02 Å. This may suggest an increase in actual titanium ratio against aluminum with increasing *n* in the generated $22(n+2)n(3n)$ crystals, because the ionic radius of $Ti^{4+}$ is larger than that of $Al^{3+}$.

Figure 3 shows HAADF-STEM images and electron diffraction (ED) patterns taken from the [100] direction of 22539 and 226412 crystals. FeAs layers are imaged as arrays of bright dumbbell-like contrasts, and cation sites of the perovskite block layers in-between are observed as less bright spots. Both HAADF-STEM images and ED patterns indicated a tetragonal cell with c/a of ~5.1 for 22539 and ~5.9 for 226412.

These values correspond with those estimated from the XRD analysis, which are 5.01 and 5.94 for 22539 and 226412 phases, respectively. Any stacking faults or superstructures were not found along the *c*-axis direction in either crystal. Unlike $(Fe_2As_2)(Ca_{n+1}(Sc,Ti)_nO_y)$, there were no sign of superstructure along the *a*-axis in ED patterns of both crystals. Perovskite-type layer of 22539 was found to be composed of mono-perovskite cell/ rock salt layer/double-perovskite cells. Interestingly, the order of the numbers of perovskite-type cells along the *c*-axis was always 1-2-1-2.

Temperature dependences of zero-field-cooled (ZFC) and field-cooled (FC) magnetization of $(Fe_2As_2)(Ca_4(Al_{0.67}Ti_{0.33})_2O_y)$, $(Fe_2As_2)(Ca_5(Al_{0.5}Ti_{0.5})_3O_y)$ and $(Fe_2As_2)(Ca_6(Al_{0.33}Ti_{0.67})_4O_y)$ are shown in Fig. 4. The $(Fe_2As_2)(Ca_6(Al_{0.33}Ti_{0.67})_4O_y)$ sample showed bulk superconductivity without intensive carrier doping. Superconducting transition was observed at 36 K and the superconducting volume fraction estimated from the ZFC magnetization at 2 K was much larger than 100% due to demagnetization effect of the bulk. Samples of $(Fe_2As_2)(Ca_4(Al_{0.67}Ti_{0.33})_2O_y)$ and $(Fe_2As_2)(Ca_5(Al_{0.5}Ti_{0.5})_3O_y)$ also showed superconducting transition with relatively large diamagnetism, indicating the possibility of superconductivity occurred in both phases. Since the 226$\underline{12}$ phase was not generated in the sample of $(Fe_2As_2)(Ca_4(Al_{0.67}Ti_{0.33})_2O_y)$ and its diamagnetism was smaller than the sample of $(Fe_2As_2)(Ca_5(Al_{0.5}Ti_{0.5})_3O_y)$, the 22539 is considered to be a superconducting phase. However, we could not determine whether the 22426 phase is superconducting or not at the present stage.

Figure 5 shows the temperature dependence of the resistivity for $(Fe_2As_2)(Ca_6(Al_{0.33}Ti_{0.67})_4O_y)$. A resistivity curve up to 300 K under 0 T is shown in the inset. The normal state resistivity exhibited a metallic behavior with a convexity, which is similar to those of $(Fe_2As_2)(Ca_{n+1}(Sc,Ti)_nO_y)$[10] and $(Fe_2As_2)(Ca_{n+1}(Mg,Ti)_nO_y)$ [11,12]. The superconducting transition was observed at $T_{c(onset)}$ of 39 K and zero resistivity was achieved at 25 K. Relatively wide superconducting transition ~15 K indicates poor grain coupling in the sample.

The new layered iron pnictides ($(Fe_2As_2)(Ca_{n+2}(Al,Ti)_nO_y)$) indicated that aluminum can be a constituent element for design of new layered iron pnictides. These compounds also suggested that local crystal structures at the perovskite-type blocking layer can be controlled by insertion of rock salt layers. Therefore, there are still considerable rooms for development of new iron pnictide superconductors in this system because of large varieties of cations as well as crystal structures. On the other hand, *a*-axis length of 22426 phase is almost comparable to that of FeSe[14]. The longest *a*-axis length among the layered iron pnictide is 4.133 Å for $(Fe_2As_2)(Ba_3Sc_2O_5)$[9]. This means local structure in $Fe_2As_2$ layer can be controlled largely by changing constituent elements and local structures at the perovskite-type layers.

**Conclusions**

New iron pnictide oxide superconductors with new crystal structures $(Fe_2As_2)(Ca_{n+2}(Al,Ti)_nO_y)$ [$n$ = 2,3,4 and $y$ ~3$n$] were discovered. The structures of the compounds are tetragonal with space group *P4/nmm* for 22426 and 2264<u>12</u> and *P4mm* for 22539 phases. These compounds have similar crystal structure with $(Fe_2As_2)(Ca_{n+1}(Sc,Ti)_nO_y)$ whereas these compounds have a rock salt layer in each blocking layer. The *a*-axis lengths of the compounds are around 3.8 Å, which is comparable to those of LiFeAs and FeSe. $(Fe_2As_2)(Ca_6(Al,Ti)_4O_y)$ exhibited bulk superconductivity with $T_c$ up to 36 K in magnetization and 39 K in resistivity measurement. The compounds indicate iron pnictides with perovskite-type layer have the variety in both crystal structure and local structure in $Fe_2As_2$ layers.

## Acknowledgements

Authors thank Mr. K. Ushiyama and N. Kawaguchi for their assistance of the experiments.

Authors also thank Dr. H. Eisaki and Prof. A. Iyo for fruitful discussions. This work was supported in part by the Ministry of Education, Culture, Sports, Science, and Technology (MEXT), Japan, through a Grant-in-Aid for Young Scientists (B) (No. 21750187).

**Figure captions**

Figure 1. Powder XRD patterns of (a) $(Fe_2As_2)(Ca_4(Al_{0.67}Ti_{0.33})_2O_6)$, (b) $(Fe_2As_2)(Ca_5(Al_{0.5}Ti_{0.5})_3O_8)$, and (c) $(Fe_2As_2)(Ca_6(Al_{0.33}Ti_{0.67})_4O_{11})$.

Figure 2. Crystal structures of (a) $(Fe_2As_2)(Ca_4(Al,Ti)_2O_6)$, (b) $(Fe_2As_2)(Ca_5(Al,Ti)_3O_9)$, and (c) $(Fe_2As_2)(Ca_6(Al,Ti)_4O_{12})$.

Figure 3. HAADF-STEM images and corresponding ED patterns of (a) $(Fe_2As_2)(Ca_5(Al_{0.5}Ti_{0.5})_3O_y)$ (space group *P*4*mm*) and (b) $(Fe_2As_2)(Ca_6(Al_{0.33}Ti_{0.67})_4O_y)$ (space group *P*4/*nmm*) crystals viewed from the [100] direction. Gray, violet, blue, and yellow circles of the inset indicate the iron, arsenic, calcium, and aluminum/titanium cation columns, respectively.

Figure 4. Temperature dependence of ZFC and FC magnetization curves for $(Fe_2As_2)(Ca_4(Al_{0.67}Ti_{0.33})_2O_y)$, $(Fe_2As_2)(Ca_5(Al_{0.5}Ti_{0.5})_3O_y)$ and $(Fe_2As_2)(Ca_6(Al_{0.33}Ti_{0.67})_4O_y)$

bulk samples measured under 1 Oe.

Figure 5. Temperature dependence of resistivity for bulk $(Fe_2As_2)(Ca_6(Al_{0.33}Ti_{0.67})_4O_y)$. Magnified resistivity curves are shown in the inset.

Figure 1

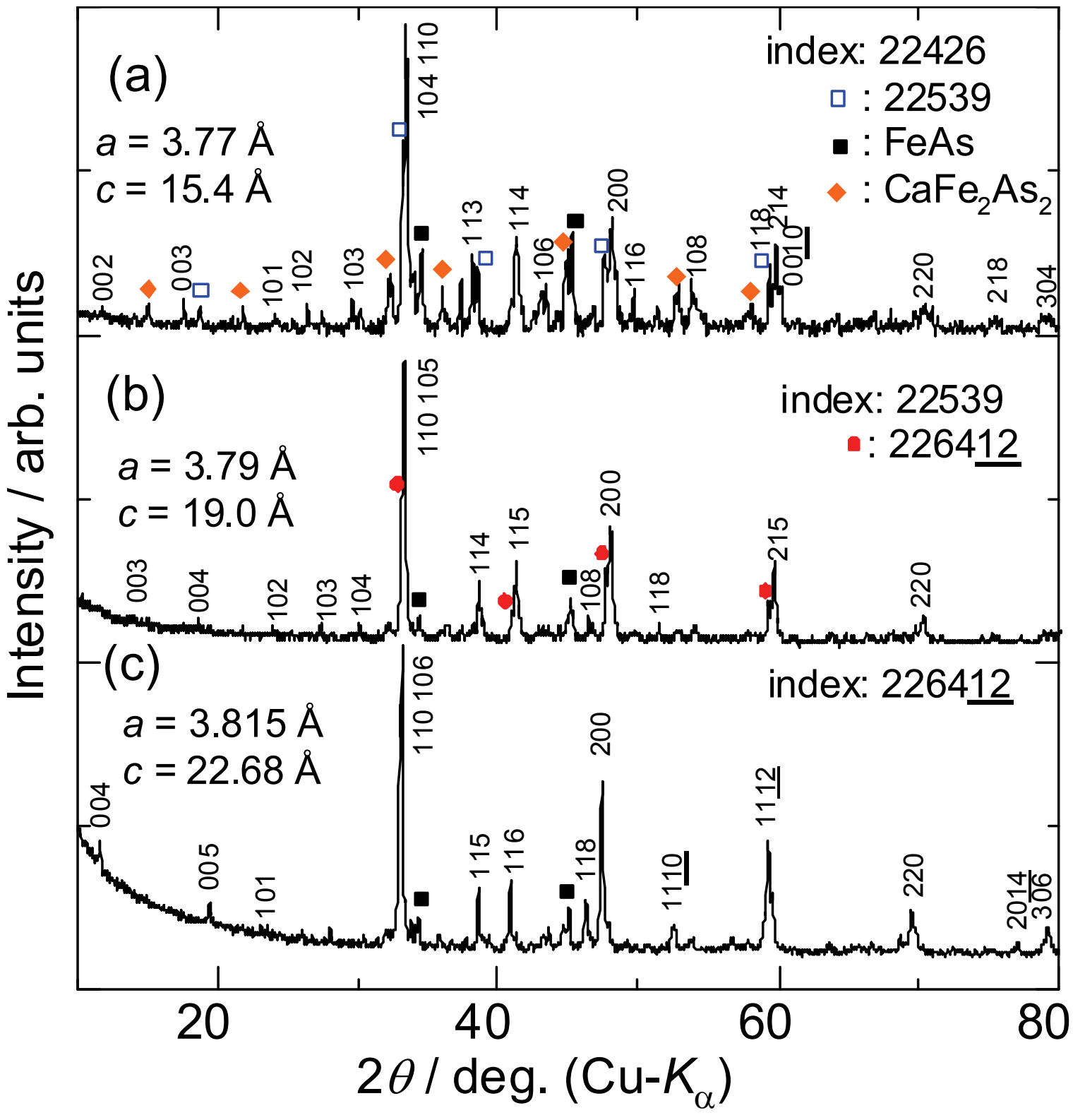

(a)
a = 3.77 Å
c = 15.4 Å
index: 22426
: 22539
: FeAs
: CaFe2As2
(b)
a = 3.79 Å
c = 19.0 Å
index: 22539
: 226412
(c)
a = 3.815 Å
c = 22.68 Å
index: 226412
Intensity / arb. units
20
40
60
80
2θ / deg. (Cu-Kα)

Figure 2

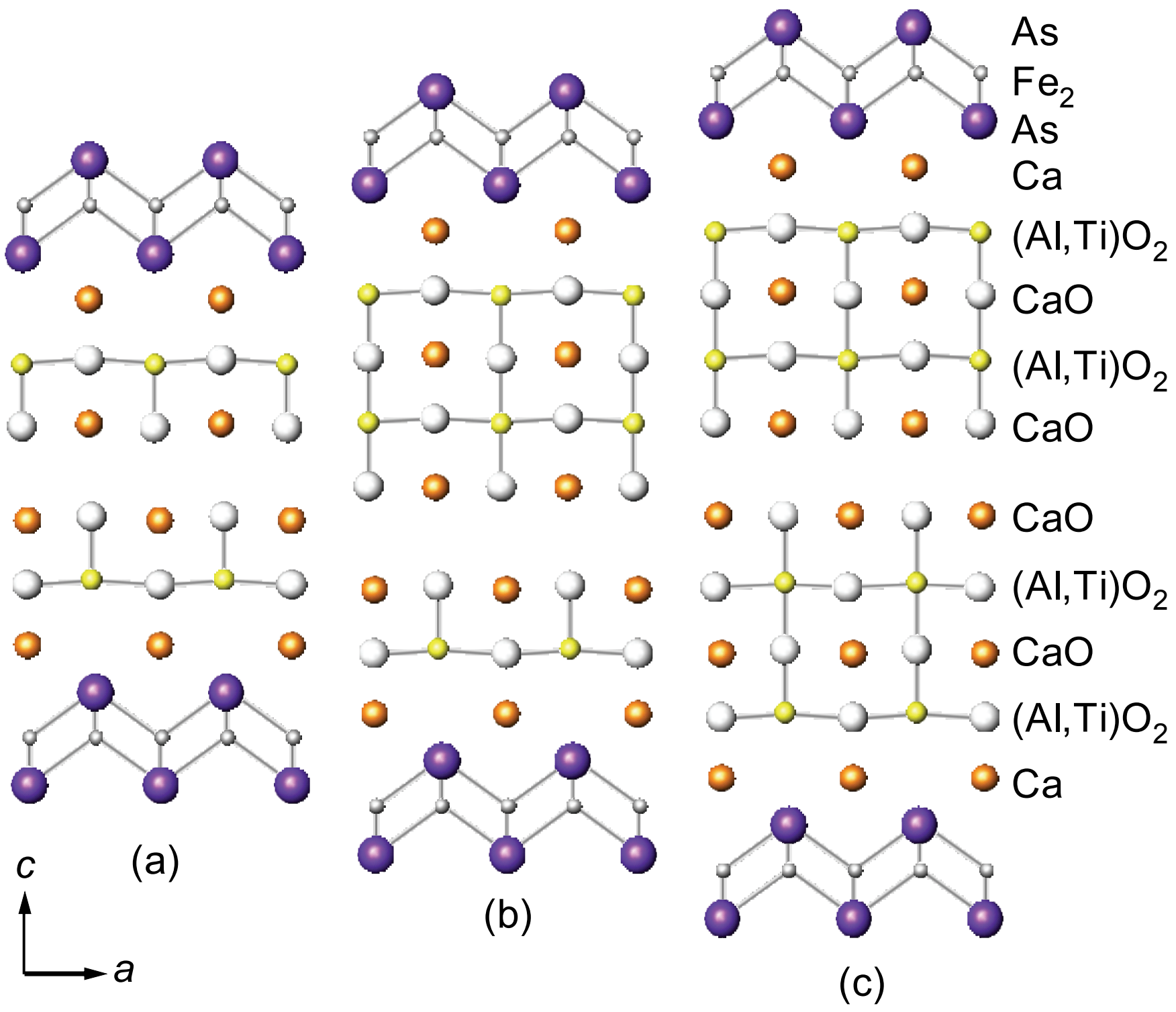

Figure 3

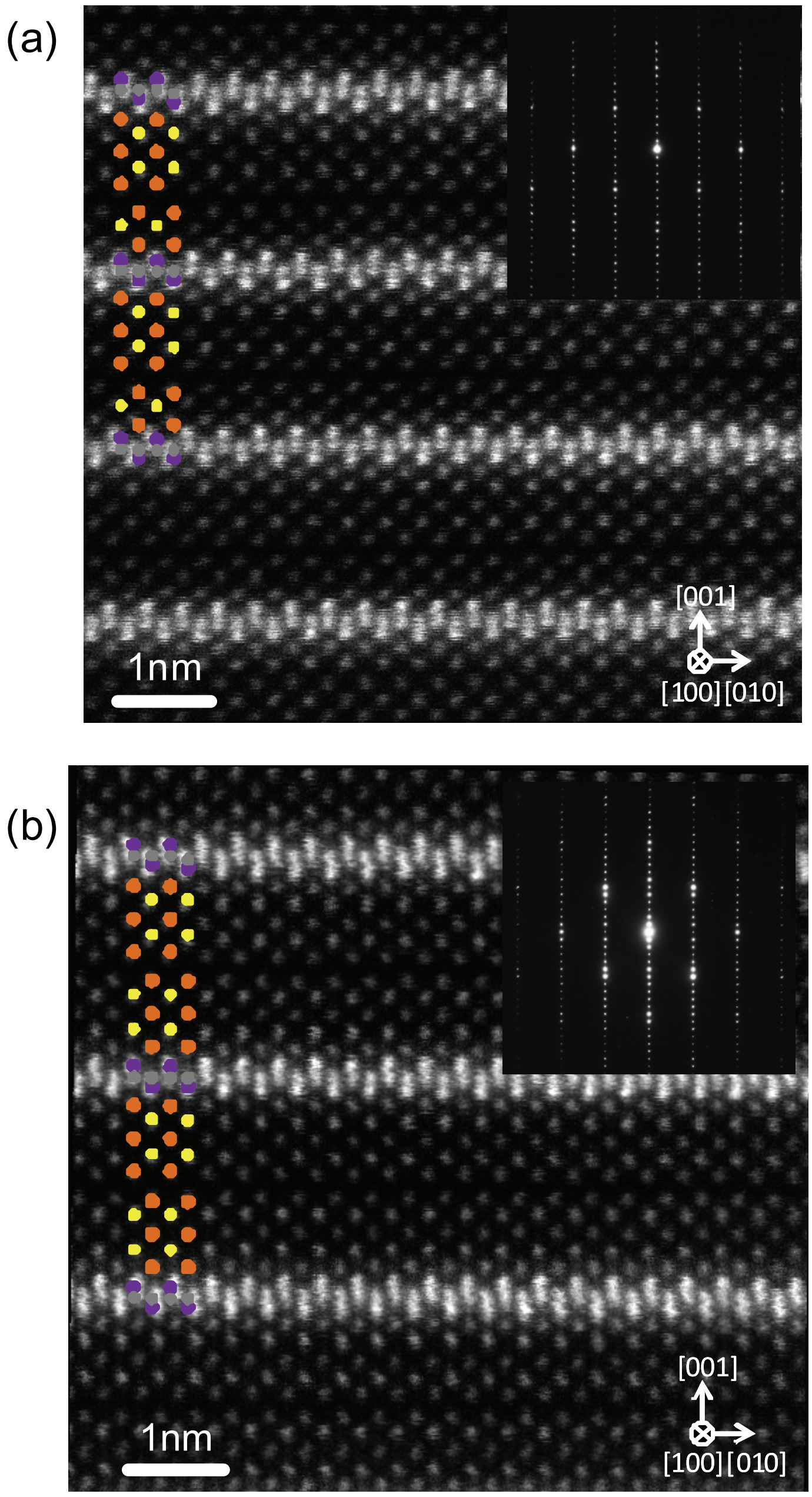

Figure 4

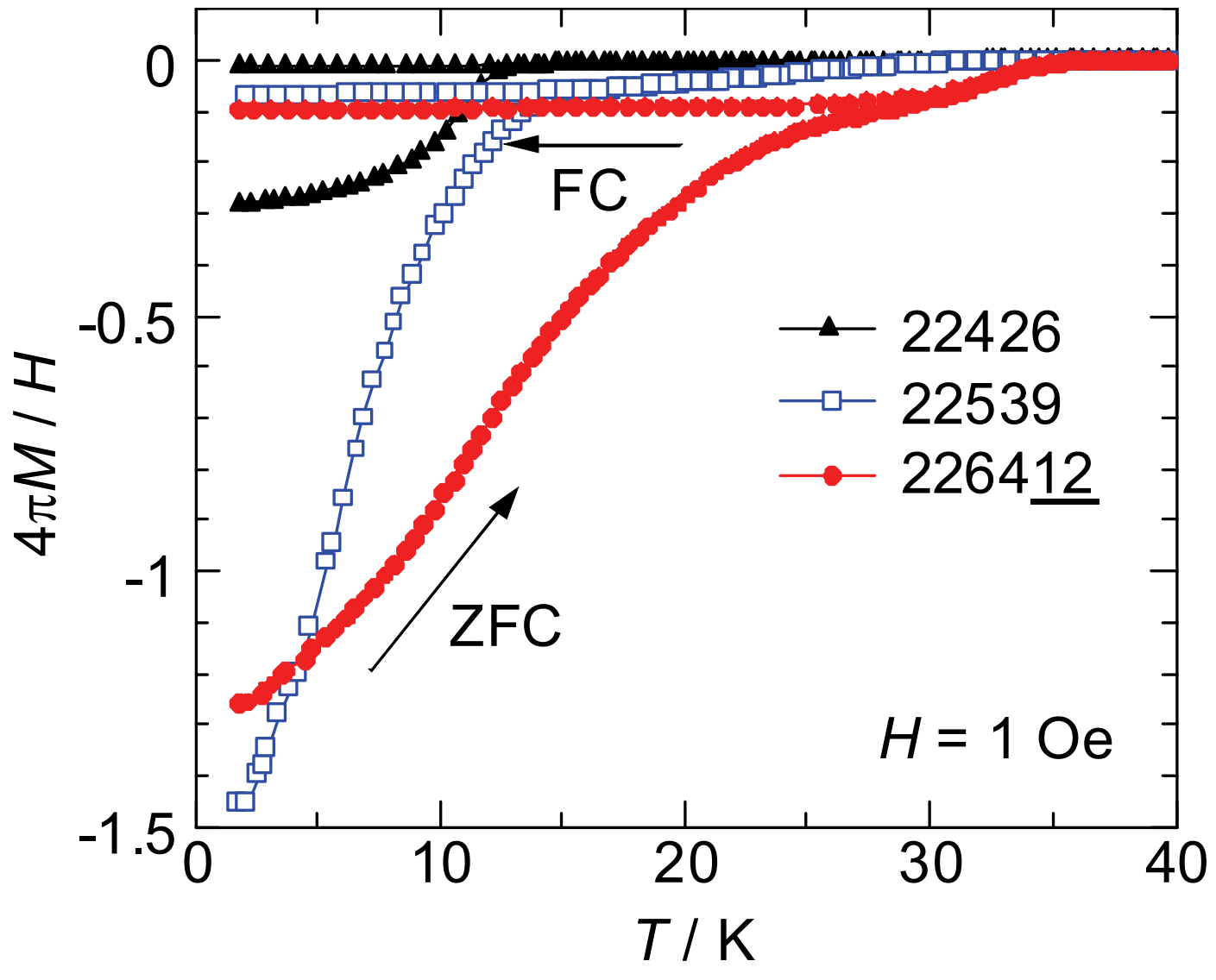

Figure 5

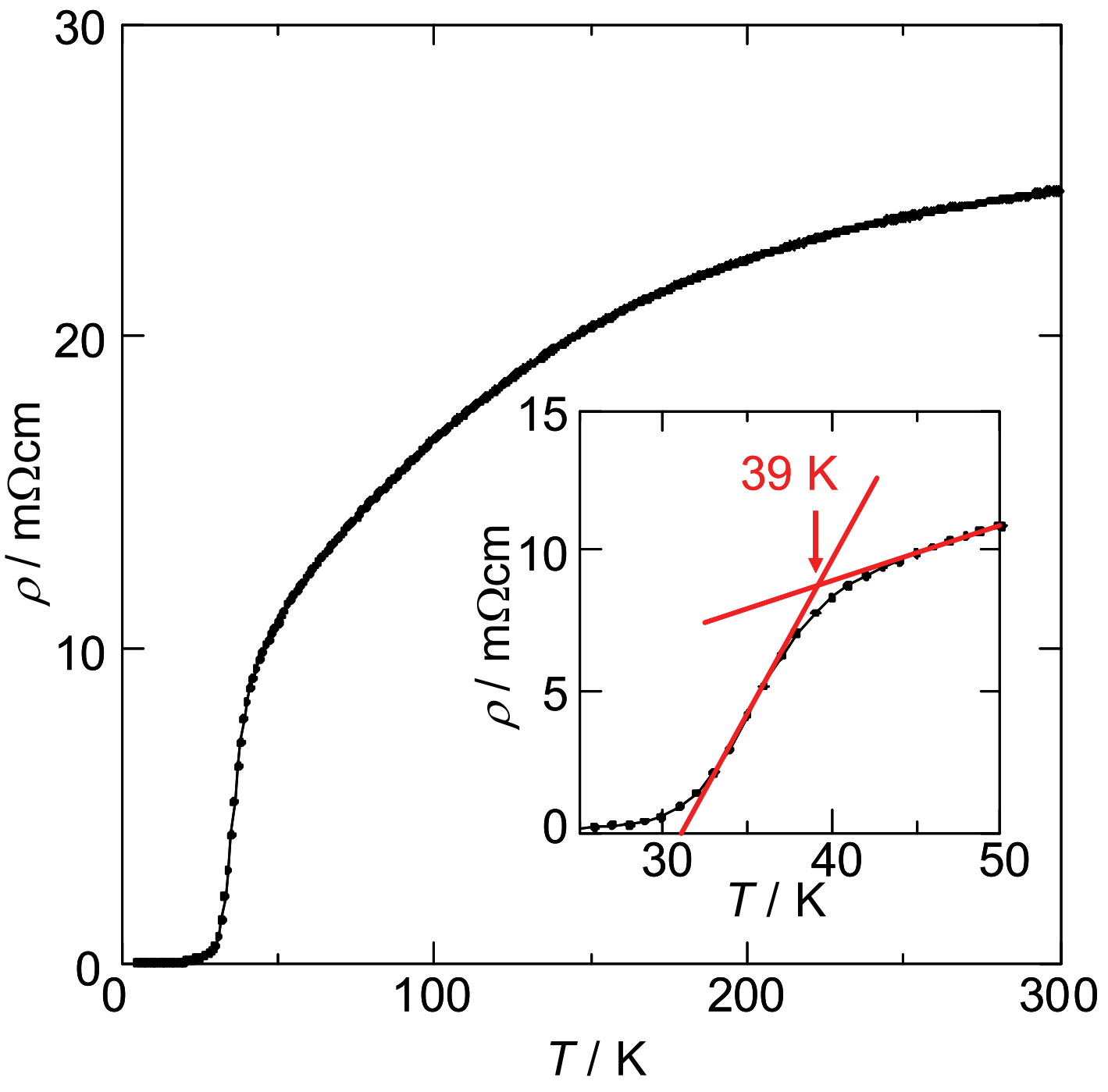